\documentclass[aps,prd,preprint,showpacs]{revtex4}
\usepackage{graphicx}
\begin{document}
\draft
\title{POSSIBILITY OF SEARCHING FOR FOURTH GENERATION NEUTRINO AT FUTURE ep COLLIDERS}
\author{A. Senol}
\email[E-mail: senol\_a@ibu.edu.tr]{}
\author{A.T. Alan}
\email[E-mail: alan\_a@ibu.edu.tr]{}
\address{Department of Physics, Abant Izzet Baysal University,
14280, Bolu, Turkey}


\begin{abstract}
We investigate the production of fourth generation neutrino in the
context of new $e\,\nu_4\,W$ magnetic dipole moment type
interaction in $ep$ collisions at the future lepton-hadron colliders. We
have obtained the mass limits of 700 GeV for THERA ($\sqrt S$=1
TeV) and 2.8 TeV for LC$\otimes$ LHC ($\sqrt S$=3.74 TeV).

\end{abstract}
\maketitle
\section{Introduction}
So far no clear experimental evidence of new physical phenomena
beyond the Standard Model (SM) has been observed. The existence of
the new massive neutrino at high energy colliders would be strong
evidence of new physics. Therefore, searching for a new massive
neutrino has high priority at collider experiments. These
neutrinos play also the basic role in cosmology for solving the
dark matter puzzle. Heavy neutrino searches at lepton-hadron
colliders has some advantages, larger masses, for instance, are
accessible compared to current $e^+e^-$ colliders
\cite{Ahmed:1994yi}.

The purpose of this paper is to investigate the possible single
production of a heavy neutrino ($\nu_4$) in $ep$ collisions in the
context of new $e\nu_4 W$ anomalous magnetic dipole moment type
interaction and analyze the background using optimal cuts. The
THERA ($\sqrt S$=1 TeV) and LC$\otimes$LHC ($\sqrt S$=3.74 TeV)
are two proposed $ep$ machine options
\cite{Abramowicz:2001qt,Sultansoy:2003wv} for which we present the
numerical results of these study.

Bounds on the possible heavy neutrino masses are obtained either
from experiments at accelerators or cosmological arguments. Among
the several arguments, the experimental data of LEP and SLAC rule
out the possibility of $\nu_4$ with mass smaller than 45 GeV
\cite{Abachi:1995de} while the cosmological limit for the upper
bound is 3 TeV \cite{Fargion:1995qb}. We assumed arrange of 200
Gev - 3 TeV for the single heavy lepton masses.

\section{Cross Section and Distributions}
\begin{figure}
  \includegraphics[width=12cm]{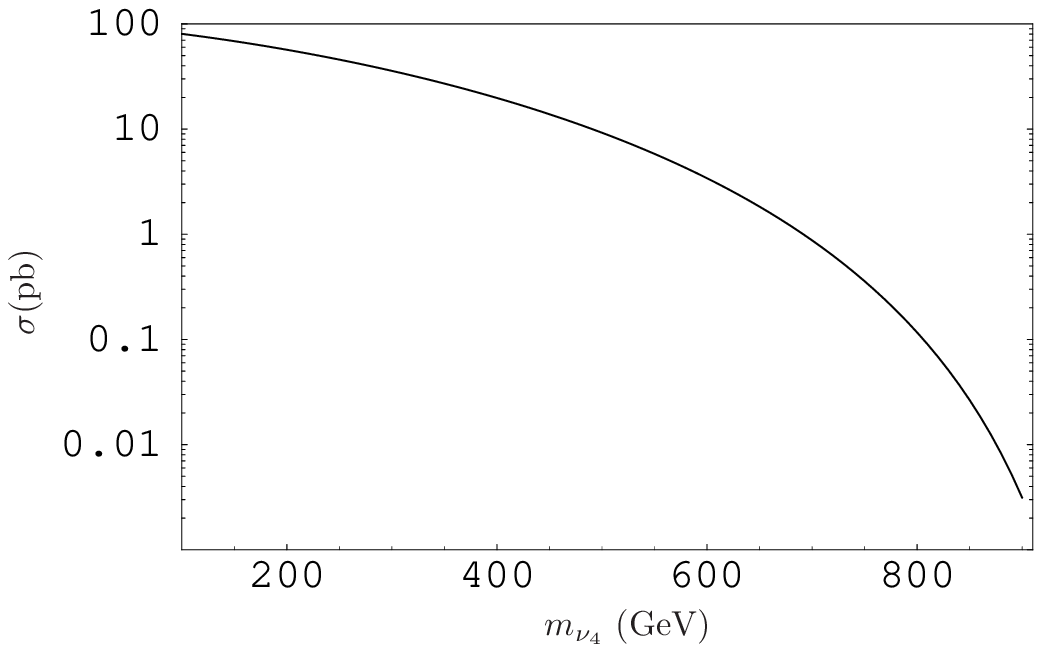}\\
\caption{The total production cross section of the subprocess
$eu\rightarrow \nu_4d$ as a function of $m_{\nu_4}$ at
THERA.}\label{fig1}
\end{figure}

\begin{figure}
\includegraphics[width=12cm]{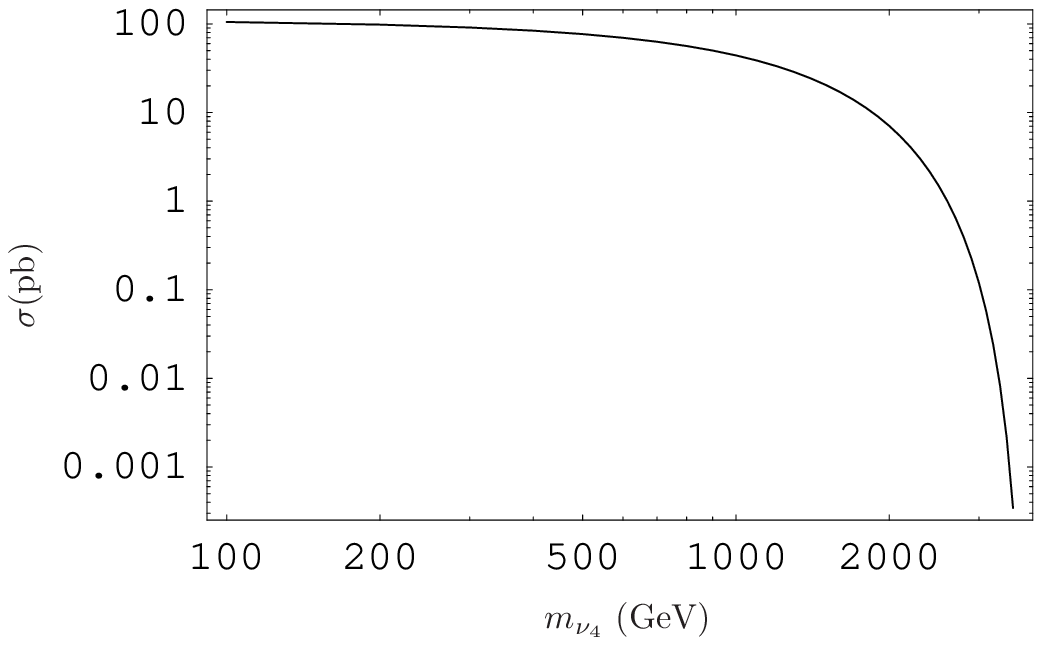}\\
\caption{The total production cross section of the subprocess
$eu\rightarrow \nu_4d$ as a function of $m_{\nu_4}$ at
LC$\otimes$LHC.}\label{fig2}
\end{figure}

The single production of heavy leptons $\nu_4$ can occur via the
t-channel $W$ exchange reaction $eq\rightarrow \nu_4 q'$, by
proposing the following new charged current $e\nu_4W$ interaction

\begin{eqnarray}
  \mathcal{L}_{cc}=\frac{g}{\sqrt 2}\,\bar e[
  \gamma_{\mu}+\frac{i}{2m_{\nu_4}}\kappa_{\nu_4}\sigma_{\mu\nu}q^{\nu}]P_L\,\nu_4\,W^{\mu}
  +h.c.,
\end{eqnarray}
which parallels the coupling structure for $\tau \nu W$ vertex
\cite{Rizzo:1997we}. In Eq.~1 $\kappa_{\nu_4}$ is the anomalous
magnetic dipole moment factor, $q$ is the momentum carried by
exchanged charged $W$ boson, $g$ denotes the gauge coupling
relative to SU(2) symmetries and $W^{\mu}$ is the field of $W$.
The differential cross section can be written as
\begin{widetext}
\begin{eqnarray}
  \frac{d\hat{\sigma}}{d\hat t}=\frac{g^4[\kappa^2_{\nu_4}((2\hat s+\hat
  t)m_{\nu_4}^2-\hat s(\hat t+\hat s)-m_{\nu_4}^4)\hat t + 4m_{\nu_4}^2\hat s(\hat s-m_{\nu_4}^2)]}{256\pi \hat s^2m_{\nu_4}^2[(\hat
  t-M^2_W)^2+\Gamma_W^2M_W^2]}
\end{eqnarray}
\end{widetext}
where $\Gamma_W$ and $M_W$ are the decay width and mass of
mediator $W$. The total production cross section is obtained by
the integration over the parton distributions in the proton:
\begin{eqnarray}
\sigma_{tot}=\int_{x_{min}}^{1}f_q(x)dx\int_{t_-}^{t_+}\frac{d\hat
{\sigma}}{d\hat t}d\hat t
\end{eqnarray}
where $t_-=-(\hat s-m_{\nu_4}^2)$, $t_+=0$ and
$x_{min}=m_{\nu_4}^2/S$. For the parton distribution functions
$f_q(x)$ we have used the MRST \cite{Martin:1998sq} and taken
$\kappa_{\nu_4}$=0.1 for illustrative purposes. The results for
the cross sections are displayed in Fig.~\ref{fig1} for the mass
range from 100 to 900 GeV at THERA, and in Fig.~\ref{fig2} from
100 GeV to 3 TeV at LC$\otimes$LHC.  For the mass values larger
than 200 GeV, the cross sections are not sensitive to the
$\kappa_{\nu_4}$ values at all. When the fourth generation
neutrino is produced, it will decay via charged current
interaction $\nu_4\rightarrow l\, W$ where $l=e,\,\mu,\,\tau$ and
we concentrate only on $\nu_4\rightarrow e\,W$ decay channel in
this work. We consider the relevant background from the
semileptonic reaction $e u\rightarrow e\,W\,d$. We present
background and signal cross section with respect to invariant mass
of $eW$ satisfying $|m_{eW}-m_{\nu_4}|<$10 GeV for the mass range
$m_{\nu_4}$=100-500 GeV and $|m_{eW}-m_{\nu_4}|<$20 GeV for the
mass range 500-1000 GeV at THERA  in Table~\ref{tab1}. In
Table~\ref{tab2}, we display background and signal cross section
with similar parameters except $|m_{eW}-m_{\nu_4}|<$50 GeV for the
mass range 1-3.4 TeV for LC$\otimes$LHC. All calculations for the
background were done using the high energy package CompHEP
\cite{Pukhov:1999gg} with CTEQ6L \cite{Lai:1994bb} distribution
function.

\begin{figure}
  \includegraphics[width=12cm]{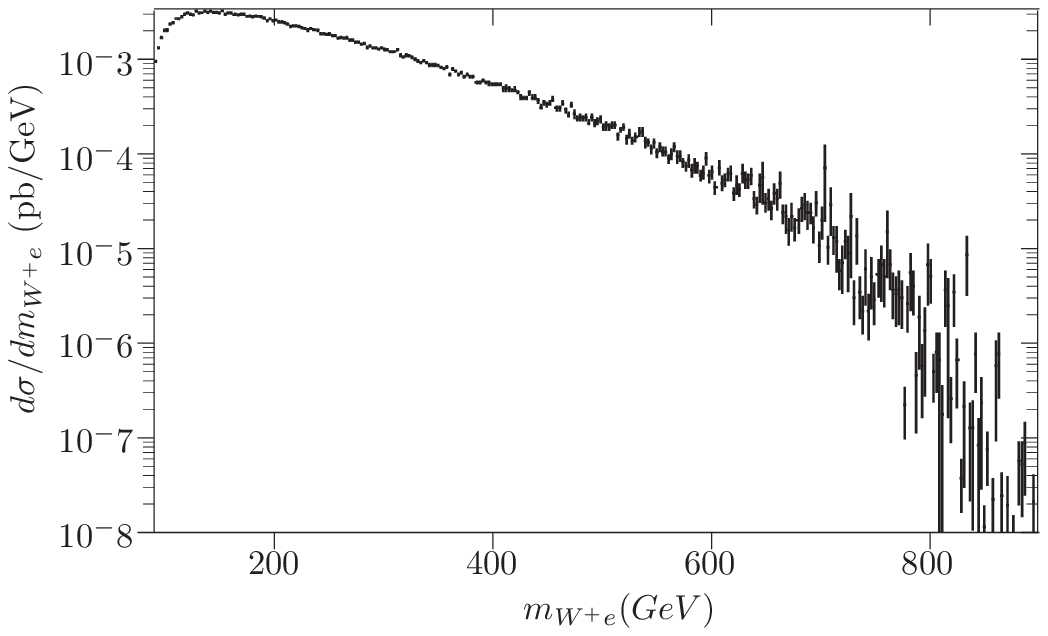}\\
\caption{The invariant mass $m_{We}$ distribution of the
background at THERA($\sqrt{S}$=1 TeV).}\label{fig3}
\end{figure}

\begin{figure}

\includegraphics[width=12cm]{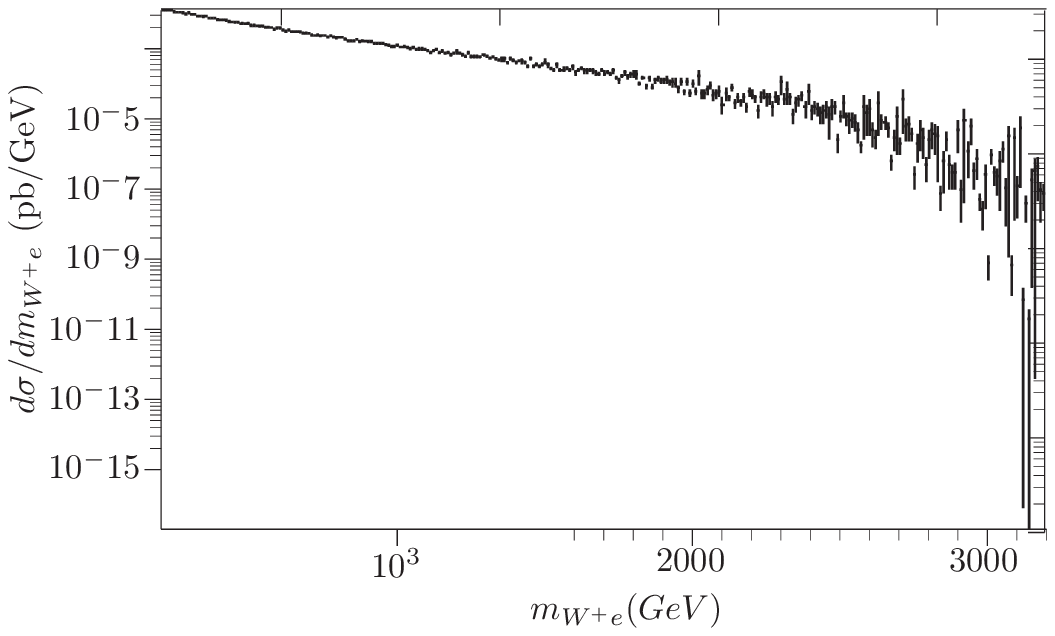}\\
\caption{The invariant mass $m_{We}$ distribution of the
background at LC$\otimes$LHC($\sqrt{S}$=3.74 TeV).}\label{fig4}
\end{figure}

\begin{figure}

\includegraphics[width=12cm]{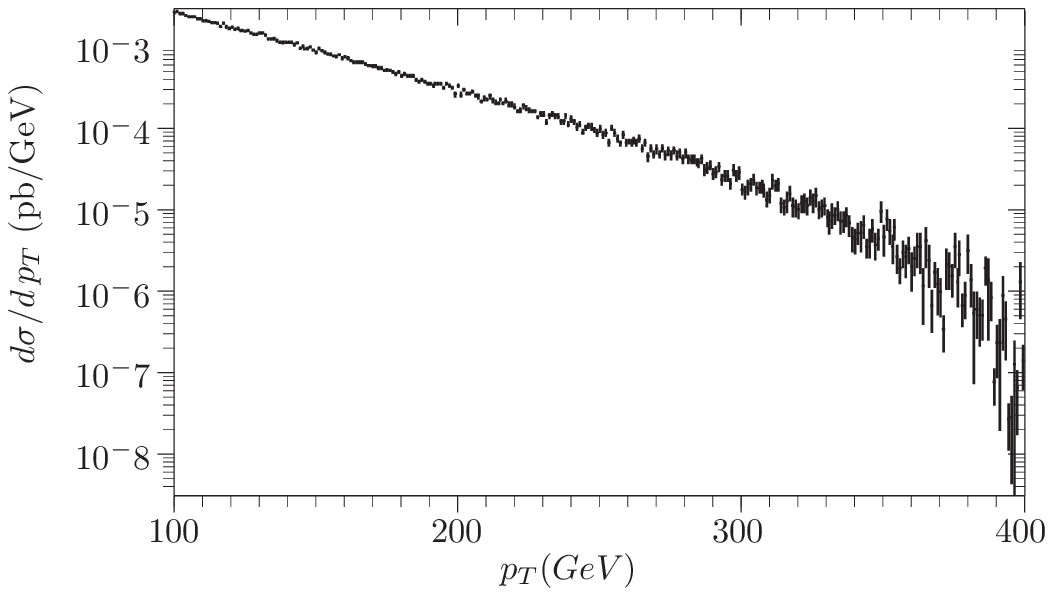}\\
\caption{$p_{T}$ distribution of the background at THERA
($\sqrt{S}$=1 TeV).}\label{fig5}
\end{figure}

\begin{figure}

\includegraphics[width=12cm]{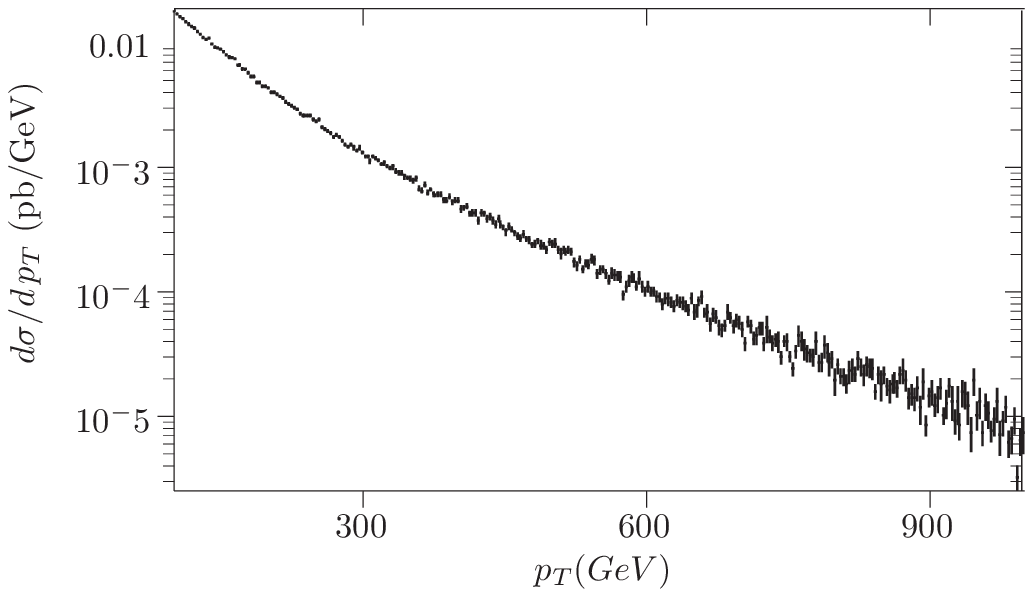}\\
\caption{$p_{T}$ distribution of the background at
LC$\otimes$LHC($\sqrt{S}$=3.74 TeV).}\label{fig6}
\end{figure}

We present the distribution of invariant mass $m_{We}$ with cuts
$p^{e^-,j}_T>10$ GeV for the background process $eu\rightarrow
e^-W^+d$ for THERA in Fig.~\ref{fig3} and for LC$\otimes$LHC in
Fig.~\ref{fig4}. Finally, in Fig.~\ref{fig5} and Fig.~\ref{fig6}
we show the $p_T$ distributions due to the $\nu_4$ production at
THERA and LC$\otimes$LHC, respectively.

\begin{table}
\centering
\begin{tabular}{l|c c c c c c c c c}
\hline\hline
 $m_{\nu_4}$(GeV)&100&200&300&400&500&600&700&800&900  \\\hline

$\sigma_S$(pb) &80.41&56.80&35.77&19.77&9.23&3.41&0.88&0.12&0.003\\
$\sigma_B$ (pb)&0.019&0.024&0.012&0.0052&0.0020&0.0013&3.2\,10$^{-4}$&4.34\,10$^{-5}$&1.31\,${10}^{-6}$ \\
  \hline\hline
\end{tabular}
\caption{Total cross sections of signal and background at THERA
for $\kappa_{\nu_4}$=0.1.}\label{tab1}
\end{table}

\begin{table}
\centering
\begin{tabular}{l|c c c c c c c c c}
\hline\hline
 $m_{\nu_4}$(GeV)& 200 & 600&1000&1400&1800&2200&2600&3000&3400  \\\hline

$\sigma_S$ (pb) &98.23&69.78&44.16&24.40&11.31&4.09&1.00&
  0.12&0.0022\\
$\sigma_B$ (pb)&0.084&0.038&0.012&0.012&0.0045&0.0014&0.00036&0.000049&1.42\,10$^{-6}$\\
  \hline\hline
\end{tabular}
\caption{Total cross sections of signal and background at LC
$\otimes$LHC for $\kappa_{\nu_4}$=0.1.} \label{tab2}
\end{table}

\section{Conclusion}
We have shown that, if we take the limit values $S/\sqrt{S+B}>5$,
without being depended on anomalous coupling for masses greater
than 200 GeV, calculated cross sections provide new 700 GeV
neutrinos at the $\sqrt{S}$=1 TeV THERA $ep$ collider, while this
limit could be as high as 2.8 TeV at the $\sqrt{S}$=3.74 TeV
LC$\otimes$LHC option. The number of events corresponding to these
limits are about 35 and 40 respectively.
\begin{acknowledgements}
This work is partially supported by Abant Izzet Baysal University
Research Found.
\end{acknowledgements}

\begin{thebibliography}{99}

\bibitem{Ahmed:1994yi}
  T.~Ahmed {\it et al.}  [H1 Collaboration],
  Phys.\ Lett.\ B {\bf 340}, 205 (1994).

\bibitem{Abramowicz:2001qt}
  H.~Abramowicz {\it et al.}  [TESLA-N Study Group],
DESY-01-011.

\bibitem{Sultansoy:2003wv}
  S.~Sultansoy,
  Eur.\ Phys.\ J.\ C {\bf 33}, S1064 (2004).

\bibitem{Abachi:1995de}
  S.~Abachi {\it et al.}  [D0 Collaboration],
FERMILAB-CONF-95-254-E.

\bibitem{Fargion:1995qb}
  D.~Fargion, M.~Y.~Khlopov, R.~V.~Konoplich and R.~Mignani,
  Phys.\ Rev.\ D {\bf 54}, 4684 (1996).



\bibitem{Rizzo:1997we}
  T.~G.~Rizzo,
  Phys.\ Rev.\ D {\bf 56}, 3074 (1997).


\bibitem{Martin:1998sq}
  A.~D.~Martin, R.~G.~Roberts, W.~J.~Stirling and R.~S.~Thorne,
  Eur.\ Phys.\ J.\ C {\bf 4}, 463 (1998).

\bibitem{Pukhov:1999gg}
  A.~Pukhov {\it et al.},
  arXiv:hep-ph/9908288.

\bibitem{Lai:1994bb}
  H.~L.~Lai {\it et al.},
  Phys.\ Rev.\ D {\bf 51}, 4763 (1995).

\end{thebibliography}
\end{document}